\documentclass[longauth]{aa}  
\usepackage{threeparttable}
\usepackage{graphicx}
\usepackage{longtable}
\usepackage{multirow}
\usepackage{natbib}
\usepackage{txfonts}
\usepackage{fixltx2e}
\usepackage{wasysym}
\usepackage{color}
\usepackage{footnote}
\bibpunct{(}{)}{;}{a}{}{,}
\begin{document}
\renewcommand{\topfraction}{0.85}
\renewcommand{\bottomfraction}{0.7}
\renewcommand{\textfraction}{0.15}
\renewcommand{\floatpagefraction}{0.66}

\newcommand{\gammaray}{$\gamma$-ray}
\newcommand{\gammarays}{$\gamma$-rays}
\newcommand{\fluxunits}{cm$^{-2}$s$^{-1}$}

\def\sig{\hbox{$\sigma$}}\def\snr{SNR\,G330.2$+$1.0}
\def\name{G330.2$+$1.0}
\def\g19{SNR\,G1.9$+$0.3}
\def\nameg19{G1.9$+$0.3}
\def\flux{\textrm{TeV}^{-1}\textrm{cm}^{-2}\textrm{s}^{-1}}
\def\snrthree{SNR G\,330.2$+$1.0}
\def\gthree{G\,330.2$+$1.0}
\def\snrone{SNR G\,1.9$+$0.3}
\def\gone{G\,1.9$+$0.3}
\def\hess{H.E.S.S.}
\def\chandra{\emph{Chandra}}
\def\xmm{\emph{XMM-Newton}}
\def\asca{\emph{ASCA}}
\def\hone{H\,I}
\def\htwo{H\,II}

\def\psrb{PSR B1259-63/LS 2883}
\def\pulsar{PSR B1259-63}
\def\Hillas{\emph{Hillas}}
\def\model{\emph{model}}
\renewcommand{\deg}{\mbox{\ensuremath{^\circ}}} 
\renewcommand{\arcmin}{\mbox{\ensuremath{^\prime}}} 
\renewcommand{\arcsec}{\mbox{\ensuremath{^{\prime \prime}}}} 
\newcommand{\beq}{\begin{equation}}
\newcommand{\enq}{\end{equation}}

\newcommand{\update}[1]{\textcolor{red}{\textbf{\boldmath#1\unboldmath}}}

\title{H.E.S.S. Observations of the Binary System \psrb\ around the 2010/2011 Periastron Passage}

\author{H.E.S.S. Collaboration
\and A.~Abramowski \inst{1}
\and F.~Acero \inst{2}
\and F.~Aharonian \inst{3,4,5}
\and A.G.~Akhperjanian \inst{6,5}
\and G.~Anton \inst{7}
\and S.~Balenderan \inst{8}
\and A.~Balzer \inst{9,10}
\and A.~Barnacka \inst{11,12}
\and Y.~Becherini \inst{13,14}
% ...
}

\institute{
Universit\"at Hamburg, Institut f\"ur Experimentalphysik, Luruper Chaussee 149, D 22761 Hamburg, Germany \and
Laboratoire Univers et Particules de Montpellier, Universit\'e Montpellier 2, CNRS/IN2P3,  CC 72, Place Eug\`ene Bataillon, F-34095 Montpellier Cedex 5, France \and
Max-Planck-Institut f\"ur Kernphysik, P.O. Box 103980, D 69029 Heidelberg, Germany \and
Dublin Institute for Advanced Studies, 31 Fitzwilliam Place, Dublin 2, Ireland \and
National Academy of Sciences of the Republic of Armenia, Yerevan  \and
Yerevan Physics Institute, 2 Alikhanian Brothers St., 375036 Yerevan, Armenia \and
Universit\"at Erlangen-N\"urnberg, Physikalisches Institut, Erwin-Rommel-Str. 1, D 91058 Erlangen, Germany \and
University of Durham, Department of Physics, South Road, Durham DH1 3LE, U.K. \and
DESY, D-15735 Zeuthen, Germany \and
Institut f\"ur Physik und Astronomie, Universit\"at Potsdam,  Karl-Liebknecht-Strasse 24/25, D 14476 Potsdam, Germany \and
Nicolaus Copernicus Astronomical Center, ul. Bartycka 18, 00-716 Warsaw, Poland \and
CEA Saclay, DSM/Irfu, F-91191 Gif-Sur-Yvette Cedex, France \and
APC, AstroParticule et Cosmologie, Universit\'{e} Paris Diderot, CNRS/IN2P3, CEA/Irfu, Observatoire de Paris, Sorbonne Paris Cit\'{e}, 10, rue Alice Domon et L\'{e}onie Duquet, 75205 Paris Cedex 13, France,  \and
Laboratoire Leprince-Ringuet, Ecole Polytechnique, CNRS/IN2P3, F-91128 Palaiseau, France
% ...
}

\date{Received November 22, 2012; accepted January 11, 2013}

\offprints{Iurii Sushch\\
\email{yusushch@physik.hu-berlin.de}\\
Mathieu de Naurois\\
\email{denauroi@in2p3.fr}
}

\abstract
{} 
%
{
In this paper we present very high energy (VHE; $E>100$ GeV) data from 
the $\gamma$-ray binary system \psrb\ taken around its periastron passage 
on 15th of December 2010 with the High Energy Stereoscopic System (H.E.S.S.) 
of Cherenkov Telescopes. We aim to search for a possible TeV counterpart of the 
GeV flare detected by the Fermi LAT. In addition, we aim to study the current periastron 
passage in the context of previous observations taken at similar orbital phases, testing 
the repetitive behavior of the source.
}
%
{
Observations at VHE were conducted with \hess\ from 9th to 16th of January 2011. The total dataset amounts to $\sim6$ h of observing time. 
The data taken around the 2004 periastron passage were also re-analysed with the 
current analysis techniques in order to extend the energy spectrum above 3 TeV to fully 
compare observation results from 2004 and 2011.
}
%
{
The source is detected in the 2011 data at a significance level of $11.5\sigma$ revealing an averaged 
integral flux above 1 TeV of $(1.01\pm0.18_{\mathrm{stat}} \pm 0.20_{\mathrm{sys}})\times10^{-12}$ \fluxunits. 
The differential energy spectrum follows a power-law 
shape with a spectral index $\Gamma = 2.92\pm0.30_{\mathrm{stat}} 
\pm 0.20_{\mathrm{sys}}$ and a flux 
normalisation at 1 TeV of 
$N_{0} = (1.95\pm0.32_{\mathrm{stat}}\pm 0.39_{\mathrm{sys}})\times10^{-12} \flux$. 
The measured lightcurve does not show any 
evidence for variability of the source on the daily scale. 
The re-analysis of the 
2004 data yields results compatible with the published ones. The differential energy 
spectrum measured up to $\sim 10$ TeV is consistent with a power law with a spectral 
index $\Gamma = 2.81\pm0.10_{\mathrm{stat}} \pm 0.20_{\mathrm{sys}}$ and a flux normalisation 
at 1 TeV of $N_{0} = (1.29\pm0.08_{\mathrm{stat}}\pm 0.26_{\mathrm{sys}})\times10^{-12} \flux$.
}
%
{
The measured integral flux and the spectral shape of the 2011 data are compatible with the results obtained 
around previous periastron passages. The absence of variability 
in the H.E.S.S. data indicates that the GeV flare 
observed by Fermi LAT in the time period covered also by H.E.S.S. 
observations originates in a different physical scenario 
than the TeV emission. Additionaly, new results compared to 
those obtained in the observations which were performed in 
2004 at a similar orbital phase, further support the hypothesis 
of the repetitive behavior of the source.
}
 

\authorrunning{H.E.S.S.\ collaboration}
\titlerunning{H.E.S.S. Observations of \psrb\ around the 2010/2011 Periastron Passage}

\maketitle

\section{Introduction}
The class of very high energy (VHE; $E>100$ GeV) $\gamma$-ray binaries 
comprises only a handful of known objects in our Galaxy: 
LS 5039 \citep{LS_5039}, LS I +61 303 \citep{LS_61303}, 
\psrb\ \citep{psrb1259_hess05} and HESS J0632+057 \citep{hess_j0632}, 
the first binary primarily discovered at TeV energies. 
This class can be extended by two more candidates: Cygnus X-1 \citep{cygnusX}, 
a stellar-mass black hole binary detected at VHEs at the 
4.1 $\sigma$ significance level, and HESS J1018-589 \citep{hess_j1018} 
which shows a point-like component spatially coincident 
with the GeV binary 1FGL J1018.6-5856 
discovered by the Fermi LAT collaboration \citep{Fermi_catalog1}.  
Only for the source presented in this paper, \psrb, the compact companion is well 
identified as a pulsar, making it a unique object for the 
study of the interaction between pulsar and stellar winds 
and the emission mechanisms in such systems. 

\psrb\ was discovered in a high-frequency radio survey devoted to the 
detection of young, distant and short-period pulsars 
\citep{johnston1, johnston2}. 
It consists of a rapidly rotating pulsar with a 
spin period of $\simeq 48$ ms and a spin-down luminosity of 
$\simeq 8\times 10^{35}$ 
erg/s in a highly eccentric ($e = 0.87$) orbit around a massive Be star. 
The pulsar moves around the companion with a period $P_{\mathrm{orb}} = 3.4$ 
years (1237 days). 

Latest optical observations with VLT UT2 \citep{negueruela} significantly 
improved the previously known parameters of the companion star LS 2883. 
The luminosity of the star is $L_{\ast}=2.3\times10^{38}$ erg s$^{-1}$.
Because of its fast rotation, the star is oblate with an equatorial 
radius of $R_{\mathrm{eq}} = 9.7\,R_{\astrosun} $ and a polar radius 
of $R_{\mathrm{pole}} = 8.1\,R_{\astrosun}$. This leads to a strong 
gradient of the surface temperature from $T_{\mathrm{eq}} \approx 27,500 K$ 
at the equator to $T_{\mathrm{pole}} \approx 34,000 K$ at the poles. 
The mass function of the system suggests a mass of the star 
$M_{\ast} \approx 30 M_{\astrosun}$ and an orbital inclination angle 
$i_{\mathrm{orb}} \approx25^{\circ}$ for the minimal neutron star mass of 
$1.4 M_{\astrosun}$. The optical observations also 
suggest that the system is located at the same distance as the 
star association Cen OB1 at $d = (2.3\pm 0.4)$ kpc \citep{negueruela}. 
The companion Be star features an equatorial disk which 
is believed to be inclined with respect to the pulsar's orbital plane 
\citep{johnston1, melatos, negueruela} in a way that the pulsar crosses 
the disk twice in each orbit just before ($\sim 20$ days) and just 
after ($\sim 20$ days) the periastron. 

Since its discovery in 1992, \psrb\ is constantly monitored by various 
instruments at all energy bands. The source shows broadband emission 
and is visible from radio wavelengths up to the VHE regime. 
The properties of 
the radio emission differ depending on the distance between 
the pulsar and the star. Radio observations \citep{johnston99, connors2002, 
johnston2005} show that 
when the pulsar is far from the periastron the observed radio 
emission consists only of the pulsed 
component, whose intensity is almost independent 
on the orbital position. But closer to 
the periastron, starting at about $t_{\mathrm{p}}-100$ d, where 
$t_{\mathrm{p}}$ is the time of periastron, the intensity starts 
to decrease up to the complete disappearance approximately at $t_{\mathrm{p}}-20$ d. 
This is followed by an eclipse of the pulsed emission for about 
35-40 days as the pulsar is behind the disk. In contrast, a transient 
unpulsed component appears and sharply rises to a level more than 
10 times higher than the flux density 
of the pulsed emission far from the periastron. 
The unpulsed component is believed to come from 
synchrotron radiation generated in the 
shocked wind zone between the relativistic pulsar wind and 
the stellar disk outflow. After the disk crossing the unpulsed emission shows a 
slight decrease with another increase around $t_{\mathrm{p}} + 20$ d 
at the second crossing of the disk. Radio observations 
around the 2007 periastron passage showed extended unpulsed emission 
with a total projected extent of $\sim120$ AU and the peak of the emission 
clearly displaced from the binary system orbit \citep{moldon2011}. This indicates that 
a flow of synchrotron-emitting particles, which can travel far away 
from the system, can be produced in \psrb. The source 
was also monitored 
around the 2010 periastron passage. The pulsed radio emission 
was monitored with Parkes telescope revealing an eclipse of the pulsed signal 
lasting from $t_{\mathrm{p}}-16$ d to $t_{\mathrm{p}}+$15 d. 
Radio emission from \pulsar\ at frequencies between 1.1 and 10 
GHz was observed using the ATCA array in the period from 
$t_{\mathrm{p}}-31$ d to $t_{\mathrm{p}}+55$ d. The detected unpulsed  emission 
around the periastron passage showed a behaviour similar to 
the one observed during previous observations \citep{psrb1259_fermi}.

\psrb\ is very well covered by X-ray observations carried out with various 
instruments like \emph{ROSAT} \citep{cominsky94}, \emph{ASCA} \citep{kaspi95, hirayama99}, 
\emph{INTEGRAL} \citep{shaw2004} and \emph{XMM-Newton} \citep{chernyakova2006}. 
The periastron passage in 2007 was monitored at the same time 
by \emph{Suzaku}, \emph{Swift}, \emph{XMM-Newton} and \emph{Chandra} \citep{chernyakova2009}. Observations around the 2010 periastron passage were 
performed by three instruments: Swift, 
Suzaku and XMM-Newton \citep{psrb1259_fermi}. 
Observations confirmed the 1-10 keV lightcurve shape 
obtained in previous periastron observations, showing a rapid X-ray 
brightening starting at about $t_{\mathrm{p}}-25$ d with 
a subsequent decrease closer to periastron and a second increase 
of the X-ray flux after periastron \citep{psrb1259_fermi}. X-ray 
observations did not show any X-ray pulsed 
emission from the pulsar. 
Unpulsed non-thermal radiation from the source appeared to be variable 
in flux and spectral index. Similarly to radio measurements, the enhancement 
of the flux occurs shortly before and shortly after the periastron. 
Unambiguously, the enhancement of the non-thermal emission results 
from the interactions of the pulsar wind with the circumstellar disk 
close to the periastron passage. 

\psrb\ was observed by \hess\ around the periastron passages in 
2004 \citep{psrb1259_hess05} and 2007 \citep{psrb1259_hess09}, leading 
to a firm detection on both occasions. In 2004, \psrb\ was observed mostly 
after the periastron, while in 2007 mostly before it. Therefore, 
the repetitive behaviour of the source, i.e. the recurrent appearance 
of the source near periastron at the same orbital phase, 
with the same flux level 
and spectral shape of the emission, was not precisely confirmed, 
since the observations covered different orbital phases. However, the similar 
dependence of the flux on the separation distance between the pulsar and the star 
for both periastron passages provides a strong hint of the repetitive behavior \citep{kerschhaggl_2011}. 
\psrb\ was not detected in observations performed far from periastron in 2005 and 2006, 
which comprised $8.9$ h  and $7.5$ h of exposure respectively.

Observations around 2004 and 2007 periastron passages 
showed a variable behaviour of the source flux with time. 
A combined lightcurve of those two 
periastron passages reveals a hint of two 
asymmetrical peaks around periastron with a 
significant decrease of the flux at the periastron itself. 
Peaks of the TeV emission roughly coincide with the flux enhancement observed 
in other wavebands as well as with the eclipse of the pulsed radio emission, 
which indicates the position of the circumstellar disk. 
This coincidence suggests that the 
TeV emission from \psrb\ may be connected to the interaction 
of the pulsar with the disk.

The paper is organised as follows: in Section 
\ref{mwl}, multiwavelength observations at the 2010/2011 periastron 
passage are reviewed. In Section \ref{obs_and_anal}, 
the dataset and analysis techniques are described and analysis results 
are presented. Results are discussed in Section \ref{discussion} and 
summarised in Section \ref{summary}.

\section{Fermi-LAT detection of a post-periastron HE flare}
\label{mwl}
H.E.S.S. observations around the most recent periastron 
passage which took place on 15th of December 2010 were 
performed as part of an extended multiwavelength (MWL) campaign 
including also radio, optical, X-ray and, for the first time, 
high energy (HE; $E>100$ MeV) observations. This paper is dedicated 
for the study of the \hess\ results in the context of the HE observations. The 
detailed study of the MWL emission from the source will be presented in the 
joint MWL paper which is currently in preparation.

Observations of the binary system \psrb\ at HEs 
were performed using the Large Area Telescope 
(LAT) on board of \emph{Fermi}. The data taken around the periastron 
passage were analysed by two independent working groups 
\citep{psrb1259_fermi, tam}, yielding similar results for 
the flaring period (see below), although there are some discrepancies 
related to the first detection period close 
to the periastron passage. Those differences do not affect however 
the conclusions drawn in this paper. 
The source was detected close to periastron 
with a very low photon flux above 100 MeV of about 
$(1-2)\times10^{-7}$ \fluxunits. After the initial 
detection, the flux decreased and the source was below 
the detection threshold 
until 14th of January, $t_{\mathrm{p}}+30$ d, when a spectacular flare was detected 
which lasted for about 7 weeks with 
an average flux $\sim10$ times higher than the flux detected 
close to the periastron \citep{psrb1259_fermi}. 
The highest day-averaged flux during 
the flare almost reached the estimate of the spin-down luminosity 
of the pulsar, which indicates a close to 100\% efficiency of the 
conversion of the pulsar's rotational energy into $\gamma$-rays.
The HE emission around the periastron 
as a function of time significantly differs from 
the two-peak lightcurves observed in other wavebands. 
The flare is not coincident with the post-periastron 
peak in radio, X-rays, and VHE $\gamma$-rays. It is also 
much brighter in comparison to the first detection 
of GeV emission close to the periastron 
passage \citep{psrb1259_fermi, tam}. 

\section{H.E.S.S. Observations and Analysis}
\label{obs_and_anal}
\subsection{The H.E.S.S. Instrument}

\begin{figure}
\centering
\vspace{-10pt}
\resizebox{\hsize}{!}{\includegraphics{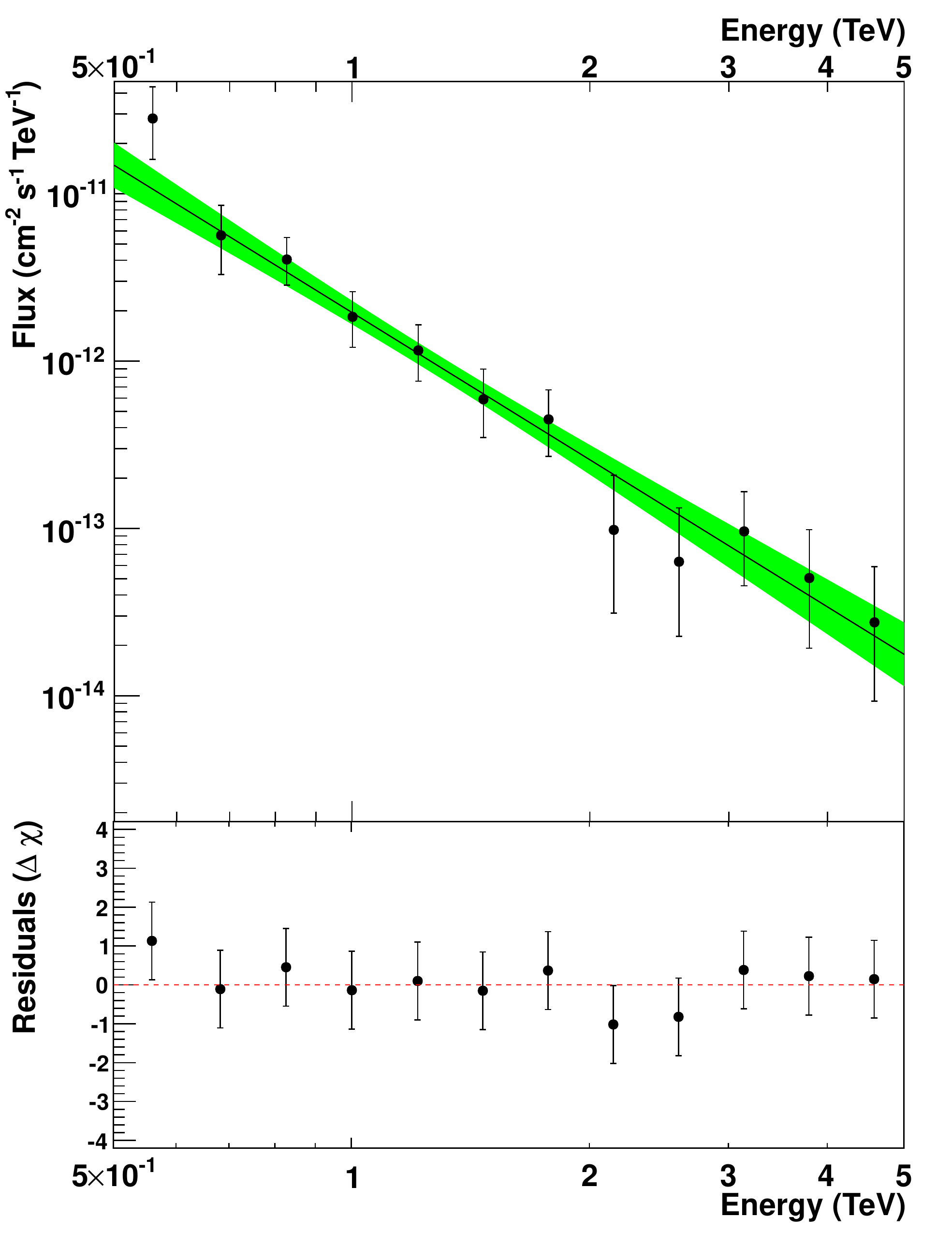}}
\caption{Overall differential energy spectrum of the VHE \gammaray\ emission from \psrb\ for the whole observation period from 9th to 16th of January 2011. The solid line denotes the spectral fit with a simple power law. 
The green band represents the 1 $\sigma$ confidence interval. Points are derived for the minimum 
significance of $1.5 \sigma$ per bin. Points' error bars represent $1 \sigma$ errors.} 
  \vspace{-10pt}
  \label{spectrum_psrb1259}
\end{figure}  

H.E.S.S. (High Energy Stereoscopic System) is an array of four 13 m diameter imaging atmospheric
Cherenkov telescopes located in the Khomas Highland, Namibia
at an altitude of 1800 m above sea level \citep{hinton_2004}. 
The telescopes are optimized for the detection of VHE $\gamma$-rays in the 
range from ~100 GeV to several tens of TeV by imaging 
Cherenkov light emitted by charged particles in 
Extensive Air Showers. The total field of view of H.E.S.S. is $5^{\circ}$. The angular 
resolution of the system is $\lesssim 0.1^{\circ}$ and the average energy resolution is about 
$15\%$. The H.E.S.S. array is capable of detecting point sources with a flux of $1\%$ 
of the Crab nebula flux at a significance level of $5\sigma$ in 25 hours 
when observing at low zenith angles \citep{crab}.

\subsection{Data Set and Analysis Techniques}
\label{analysis}
\psrb\ observations were scheduled to cover the post-periastron 
period from January to March 2011. The source was not visible for 
H.E.S.S. before and at the periastron passage. The observations 
resulted in a rather small dataset due to unfavorable weather conditions. 
The collected data correspond to 6 h of 
livetime after the standard quality selection 
procedure \citep{crab}. These data were taken in five nights, 
namely January 9/10, 10/11, 13/14, 14/15 
and 15/16. The observations were performed at a relatively 
high average zenith angle of 48\deg\ and with a mean offset angle of 0.55\deg\ from the 
test region centered at $\alpha_{\mathrm{J2000}} = 13^{\mathrm{h}}02^{\mathrm{m}}48^{\mathrm{s}}$, 
$\delta_{\mathrm{J2000}} = -63\deg50\arcmin09\arcsec$ \citep{wang_2004}.

The data were analysed using the \emph{model analysis}\footnote{ParisAnalysis 
software version 0-8-18.} technique with \emph{standard cuts} \citep{deNaurois09}. 
The test region was \emph{a priori} defined as a circle with radius 
0.1\deg\ (i.e. $\theta^2 < 0.01 ^{\circ^{2}}$, where $\theta$ is defined as 
the angular distance between the \gammaray\ event and the nominal target 
position) which is the standard size for point-like sources. 
The \emph{Reflected Region Background} technique was 
used for the background subtraction. 
The analysis results were cross-checked with an alternative analysis 
chain\footnote{H.E.S.S. Analysis Package (HAP) version 11-02-pl04.} 
using a standard Hillas reconstruction \citep{crab} method for 
$\gamma$/hadron separation and an independent calibration 
of the raw data. Both analysis chains yielded consistent results.

\begin{table*}[t]
  \small
  \centering
  \begin{threeparttable}[b]
    \caption{Analysis results of the H.E.S.S. data for the full observation period as well as for the pre-flare and flare periods. The latter two are defined by the beginning of the HE flare (see text). $N_{\mathrm{ON}}$ and $N_{\mathrm{OFF}}$ are numbers of ON and OFF events, $\alpha$ is the background normalisation and  $N_{\gamma}$ is the number of excess photon events.}
    \label{analysis_res}
    \tabcolsep=0.11cm
    \begin{tabular}{c | c c c c c c | c c c }
      \hline
      \hline
      &&&&&&&&\\
      Dataset& Livetime [h]& $N_{\mathrm{ON}}$ & $N_{\mathrm{OFF}}$ & $\alpha$ & $N_{\gamma}$ & Significance [$\sigma$]& $\Gamma$ & $N_{0}$ [$10^{-12}\,\flux$]& Flux($E>1$ TeV) \\  
      &&&&&&&&&[$10^{-12}$ \fluxunits]\\
      \hline
      &&&&&&&&&\\
      Full & 6.2& 112& 365& 0.077& 84.0&11.5&$2.92\pm0.30_{\mathrm{stat}} \pm 0.20_{\mathrm{syst}}$&$1.95\pm0.32_{\mathrm{stat}}\pm 0.39_{\mathrm{syst}}$&$1.01\pm0.18_{\mathrm{stat}} \pm 0.20_{\mathrm{syst}}$\\ 
      &&&&&&&&&\\
      Pre-flare & 2.65& 44& 133& 0.076& 33.9& 7.4& $2.94\pm0.52_{\mathrm{stat}} \pm 0.20_{\mathrm{syst}}$& $2.15\pm0.56_{\mathrm{stat}}\pm 0.43_{\mathrm{syst}}$ & $1.11\pm0.29_{\mathrm{stat}} \pm 0.22_{\mathrm{syst}}$\\
      &&&&&&&&&\\
      Flare & 3.59& 68& 232& 0.077& 50.1& 8.5&  $3.26\pm0.49_{\mathrm{stat}} \pm 0.20_{\mathrm{syst}}$& $1.81\pm0.39_{\mathrm{stat}}\pm 0.36_{\mathrm{syst}}$& $0.80\pm0.22_{\mathrm{stat}} \pm 0.16_{\mathrm{syst}}$\\
      \hline
    \end{tabular}
  \end{threeparttable}
\end{table*}

\subsection{Energy Spectrum}
The source was detected at a 11.5 $\sigma$ level \citep{lima} 
(see Table \ref{analysis_res}). 
A spectral analysis of the detected excess events shows that the differential 
energy spectrum of photons is consistent with a simple power law $\mathrm{d}N/\mathrm{d}E = 
N_{0} \left( E/1 \mathrm{TeV} \right)^{-\Gamma}$
with a flux normalisation at 1 TeV of $N_{0} = (1.95\pm0.32_{\mathrm{stat}}\pm 0.39_{\mathrm{syst}})\times10^{-12}\, \flux$ 
and a spectral index $\Gamma = 2.92\pm0.30_{\mathrm{stat}} \pm 0.20_{\mathrm{syst}}$ 
(see Fig. \ref{spectrum_psrb1259} 
and Table \ref{analysis_res}) 
with a fit probability of 0.64. The integral flux above 1 TeV averaged over the 
whole observation period is $F(E>1\mathrm{TeV})=(1.01\pm0.18_{\mathrm{stat}} 
\pm 0.20_{\mathrm{sys}})\times10^{-12}$ \fluxunits.

\subsection{Lightcurves}
\label{lightcurves}
\begin{figure}
\centering
 \resizebox{\hsize}{!}{\includegraphics{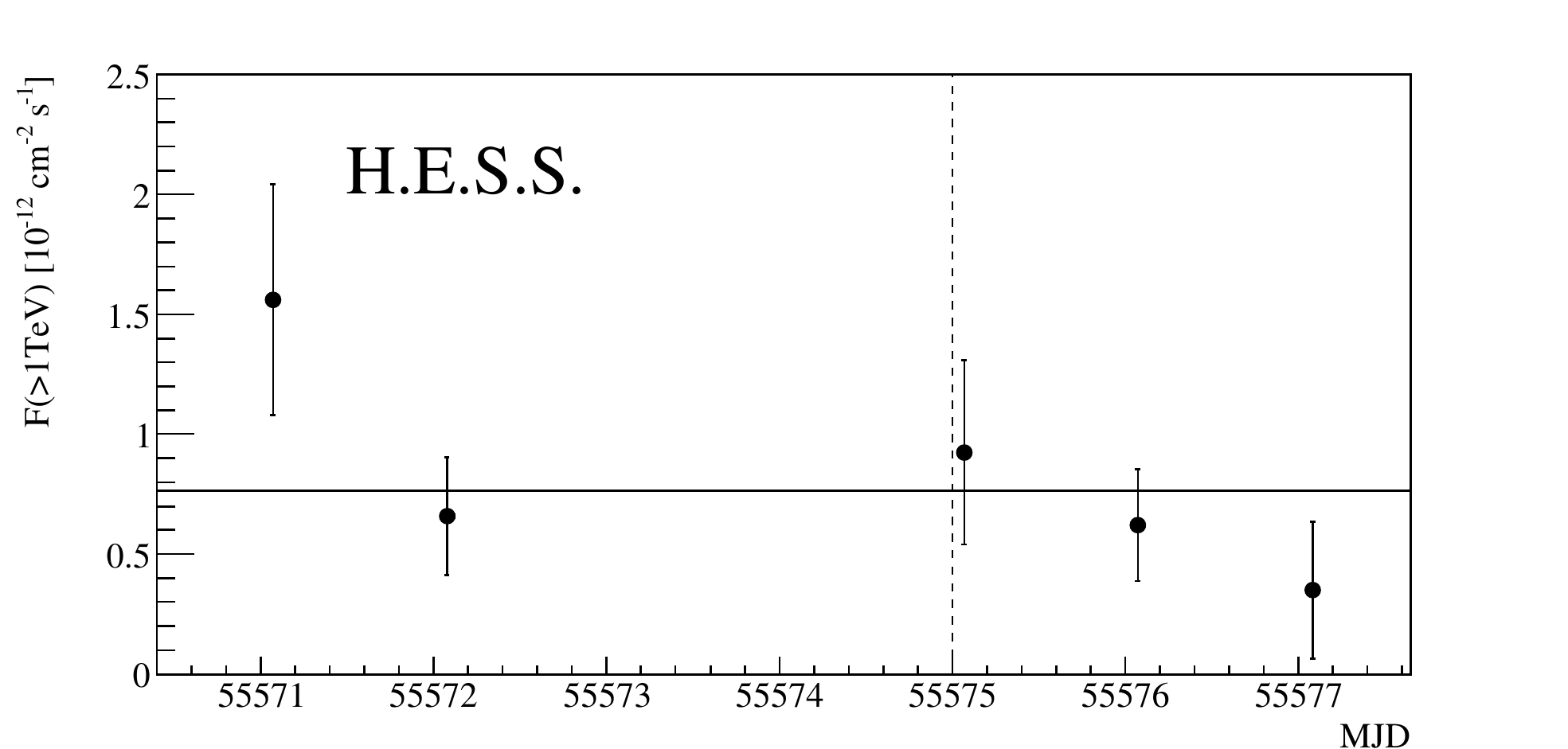}}
\caption[Lightcurve]{Integrated photon flux above 1 TeV for individual observation nights. The solid horizontal line indicates the fit of a constant to the distribution. The flare start date is indicated by the dashed vertical line.}
\label{night_by_night}
\end{figure}

In order to check for variability of the source a lightcurve has been produced on 
a night-by-night basis assuming the photon spectral index obtained 
in the spectral fit (Fig. \ref{night_by_night}). The spectral index was 
fixed at the value obtained in the spectral analysis of the total data because 
of the low statistics for each individual night. 
The lightcurve is consistent with a constant 
resulting in a mean flux of $(0.77\pm0.13)\times10^{-12}$ \fluxunits\ 
(horizontal line in Fig. \ref{night_by_night}) with $\chi^{2}/NDF=6.35/4$ (probability of 0.17) 
yielding no evidence for variability in the 7-night observation period. 
For each individual night the 
source is detected at a statistical significance level $>3\sigma$ 
except the last point which significance is only $1.5 \sigma$.

For a comparison with the GeV flare (see Section \ref{flare_search}) the whole dataset was 
divided into two datasets: "pre-flare" ($t_{\mathrm{p}}+26$ d to 
$t_{\mathrm{p}}+29$ d) and "flare" ($t_{\mathrm{p}}+30$ d to $t_{\mathrm{p}}+32$ d). 
These two datasets were analysed independently revealing similar fluxes 
and significance levels (see Table \ref{analysis_res}). A spectral 
analysis of the two datasets shows that both spectra are consistent with a simple power law, yielding 
similar values of the spectral index (see Table \ref{analysis_res}). Both spectral indices are 
consistent with the one obtained for the total dataset. These results are discussed 
in Section \ref{flare_search}.

\subsection{Re-analysis of the 2004 Data}

\begin{figure}
  \centering
  \resizebox{\hsize}{!}{\includegraphics[width=\textwidth]{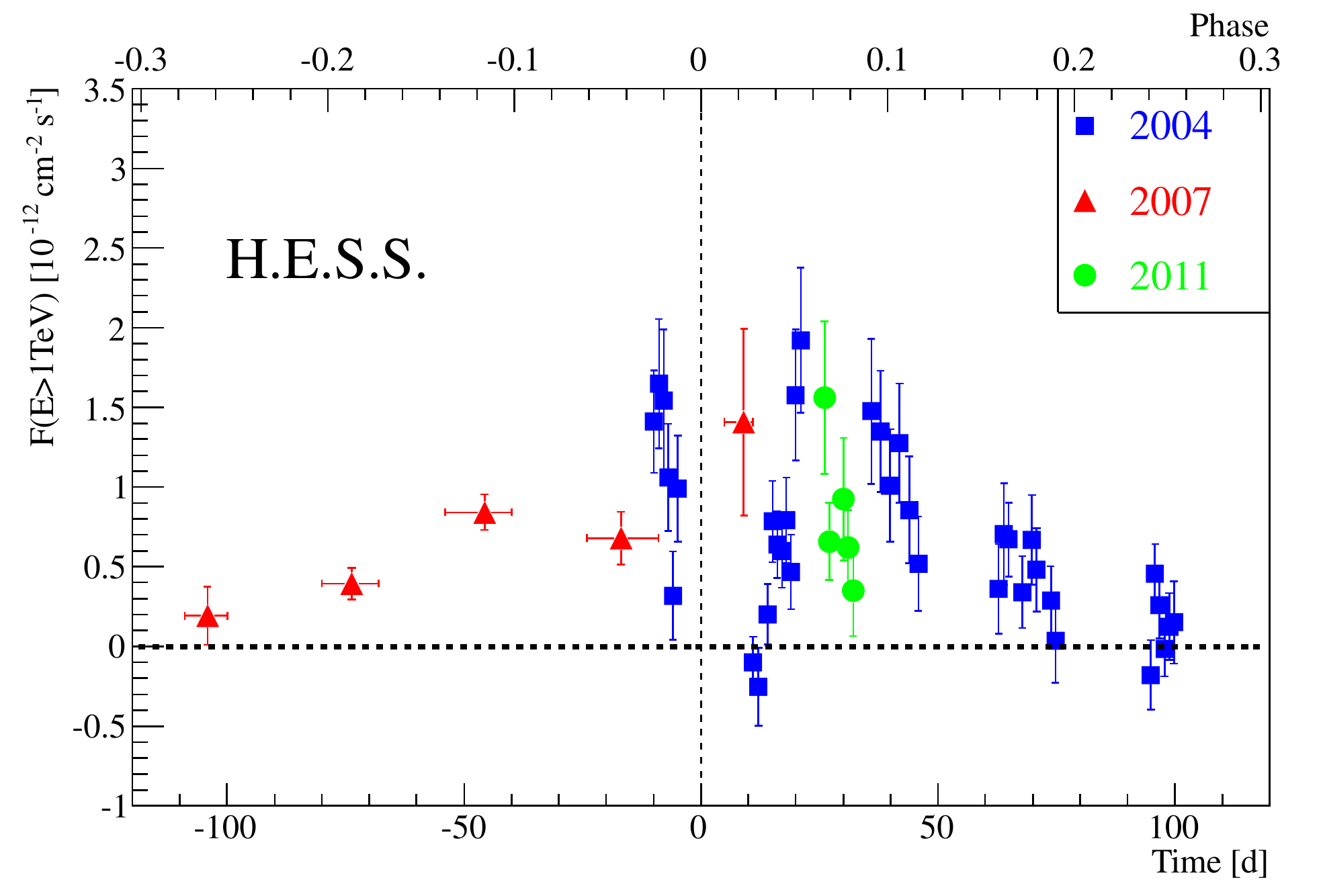}}
  \caption{Integrated photon flux above 1 TeV as a function of the time with respect to the periastron passage indicated with the vertical dashed line. The corresponding orbital phases (mean anomaly) are shown on the upper horizontal axis. The data from the 2004 (blue squares) \citep{psrb1259_hess05}, 2007 (red triangles) \citep{psrb1259_hess09} and 2011 (green circles) observation campaigns are shown. For the 2004 and 2011 data the flux is shown in daily bins while for the 2007 data the flux is shown in monthly bins for clarity.}
  \label{overall_lc}
\end{figure}  

For the data taken around the 2004 periastron passage the energy spectrum 
had been measured only up to $\sim 3$ TeV \citep{psrb1259_hess05} 
while for the much smaller dataset of 2011 observations 
the spectrum was measured up to $> 4$ TeV and for the comparable dataset of 
2007 observations up to $> 10$ TeV \citep{psrb1259_hess09} using 
more advanced analysis techniques with a better understanding of weak fluxes. 
In order to fully compare observation results around different 
periastron passages (see below), the 2004 data were 
re-analysed with the current analysis techniques, the same as used for the 
analysis of the 2011 data described above. The re-analysis results are 
compatible with the published ones. The differential energy spectrum measured 
up to $\sim 10$ TeV is consistent with a power law with a spectral index 
$\Gamma = 2.8\pm0.1_{\mathrm{stat}} \pm 0.2_{\mathrm{syst}}$ and a flux 
normalisation at 1 TeV of $N_{0} = (1.29\pm0.08_{\mathrm{stat}}\pm 
0.26_{\mathrm{syst}})\times10^{-12}\, \flux$. The new analysis 
of the 2004 data is therefore compatible with the published results when 
extrapolated above 3 TeV.


\section{Discussion}
\label{discussion}

\subsection{Comparison with Previous H.E.S.S. Observations}

In Fig. \ref{overall_lc}, the integrated photon flux above 1 TeV 
as a function of time with respect to periastron (indicated by 
the dashed vertical line) is shown. The lightcurve compiles the data from all 
three periastron observation campaigns spanning from 100 days before to 
100 days after the periastron. The observed flux from the 
2010/2011 observation campaign is compatible with the flux 
detected in 2004 at the similar orbital phases. 
Observation periods from 2004 and 2007 were separated 
in time with respect to the periastron position, i.e. observations in 2004 
were performed mainly after and in 2007 mainly before the periastron. 
Therefore, it was impossible to directly confirm the 
repetitive behaviour of the 
source by comparing observations of \psrb\ at the same orbital 
phases. In this perspective, although the 2011 observations do not 
exactly overlap with the orbital phases of previous studies, 
they cover the gap in the 2004 data post-periastron lightcurve 
and the integrated flux follows the shape of the lightcurve 
yielding a stronger evidence for the repetitive behaviour of the source.

\begin{figure}
  \centering
  \resizebox{\hsize}{!}{\includegraphics[width=\textwidth]{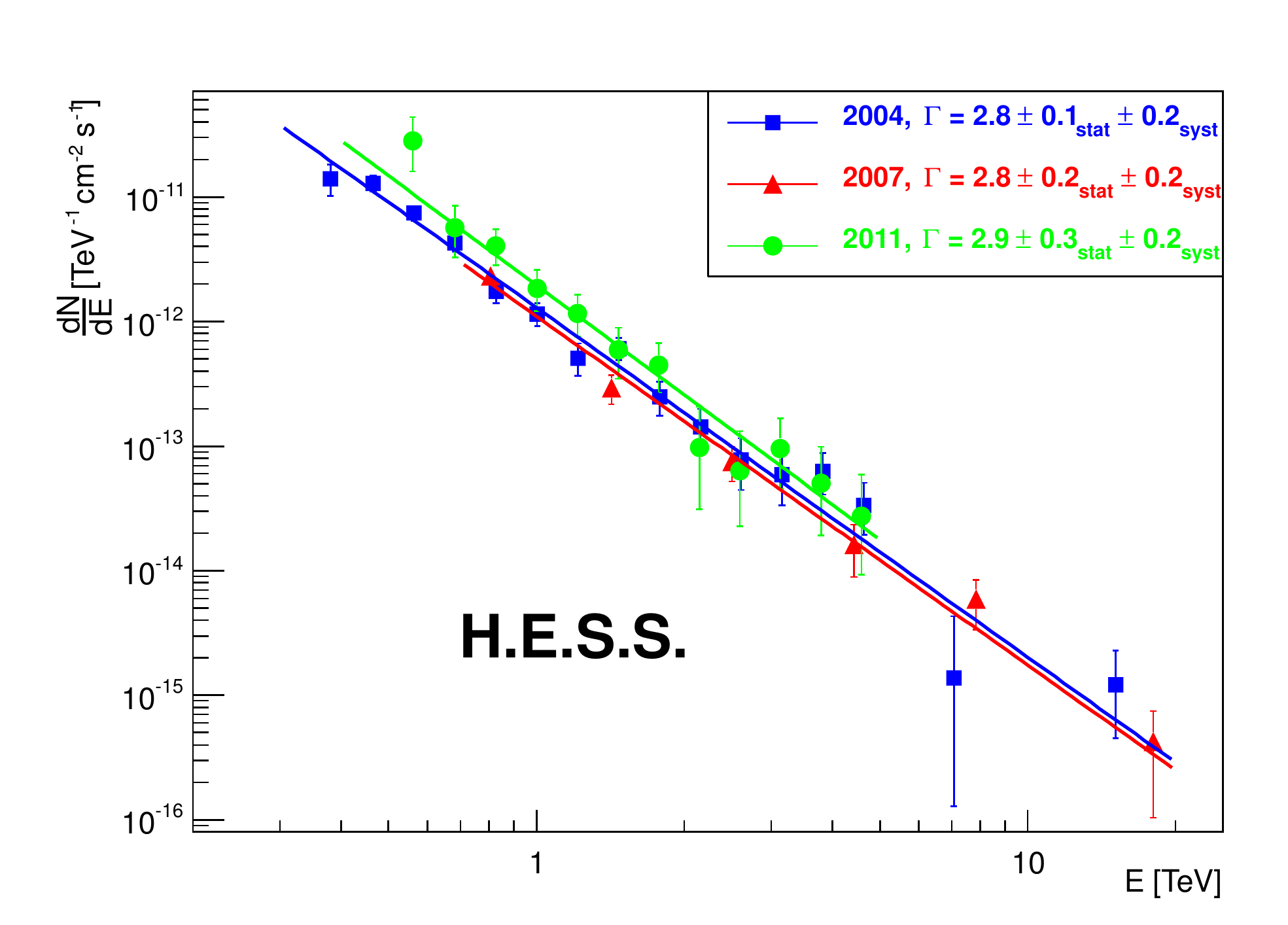}}
  \caption{Differential energy spectra of the VHE \gammaray\ emission from \psrb\ for 
the data collected around the 2004 (blue squares), 2007 (red triangles) and 2010/2011 (green circles) periastron 
passages. For the 2004 data the spectrum presented in this paper is shown. The 2007 spectrum 
is extracted from \citet{psrb1259_hess09}.}
  \label{all_spectra}
\end{figure}

The spectral shape of the VHE \gammaray\ emission from \psrb\ around 
the 2010/2011 periastron passage is similar 
to what was observed during previous periastron passages (Fig. \ref{all_spectra}). 
The photon index of $2.92\pm0.25_{\mathrm{stat}}\pm 0.2_{\mathrm{syst}}$ 
inferred from the 2011 data is well compatible 
with previous results. The spectrum measured for the 2011 data can be 
resolved only up to $\sim 4$ TeV, which is explained by a very low statistics at 
higher energies due to a short exposure of the source.

\subsection{Search for the equivalent "GeV Flare" in the H.E.S.S. data}
\label{flare_search}

The absence of the flux enhancement during the GeV flare at radio 
and X-ray wavebands indicates that the GeV flare may be created 
by different physical processes from the ones responisble 
for the emission at other wavelengths. The VHE post-periastron data 
obtained with H.E.S.S. around the 2004 periastron passage 
do not show any evidence of a flux outburst at orbital phases 
at which the GeV flare is observed. However, the H.E.S.S. observations around the 2004 
periastron passage do not comprise the orbital phase when the GeV flare starts. 
Also, in order to compare H.E.S.S. 2004 data with the GeV flare observed after the 2010 
periastron passage, one has to assume that the GeV flare is a periodic phenomenon 
which may not be the case. 
The H.E.S.S. data taken between 9th and 16th of January in 2011 provide a three-day overlap 
in time with the GeV flare. Therefore, it is possible to directly 
study any flux enhancement in the VHE band on 
the timescale of the HE flare. In order to improve the sensitivity of the 
variability search the whole period of the H.E.S.S observations was 
divided into two time periods of almost equal length: before ("pre-flare") and during 
("flare") the HE flare (see Section \ref{lightcurves}). The pre-flare and flare datasets analysis 
results are presented in Table \ref{analysis_res}.

\begin{figure}
  \centering
  \includegraphics[width = \linewidth]{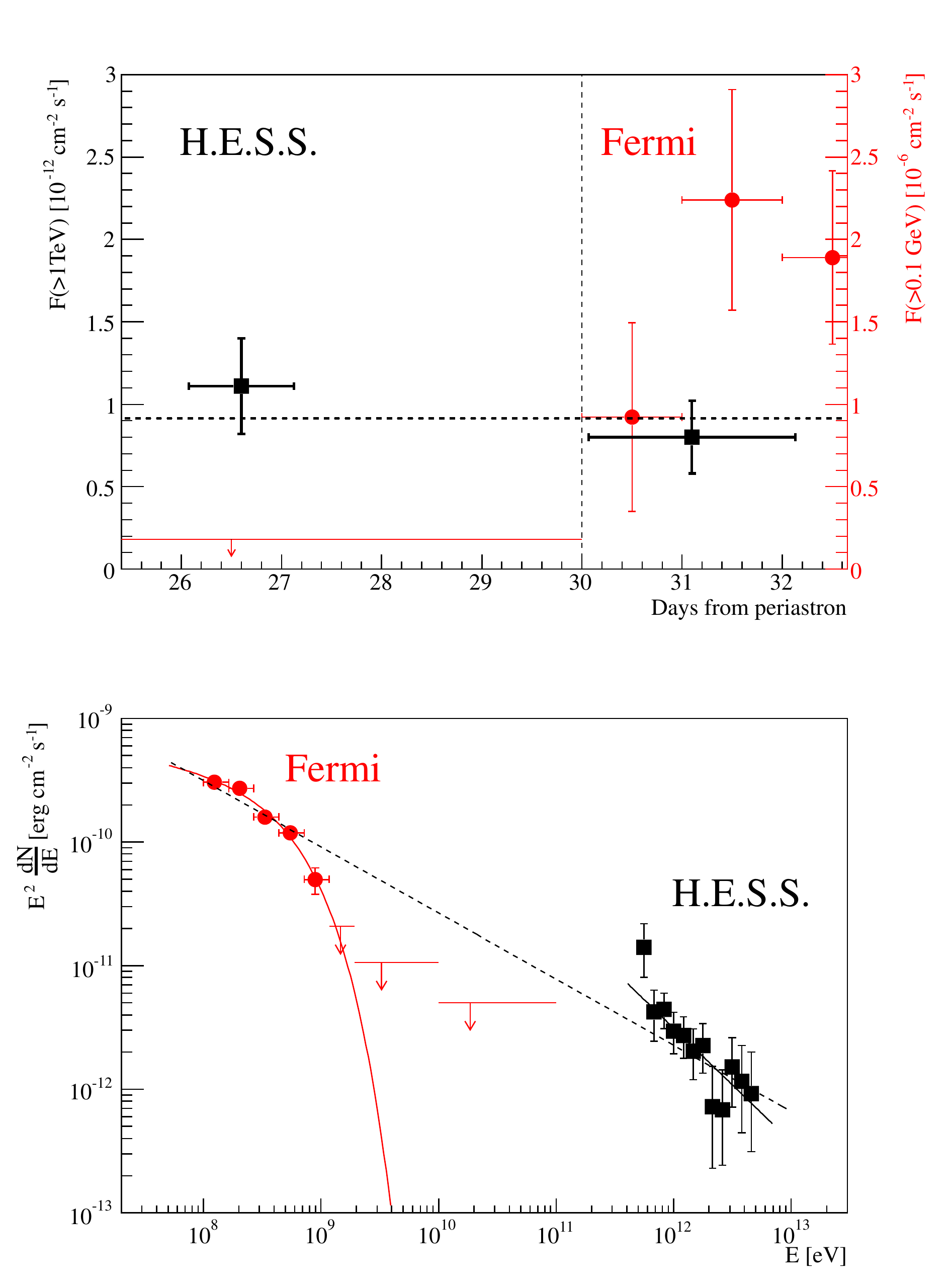}
  \caption{(Top) The integrated photon fluxes above 1 TeV for the pre-flare and flare periods (see text) are shown as black filled boxes. The dashed horizontal line shows the best fit with a constant. The HE data points above 0.1 GeV as reported by \citet{psrb1259_fermi} are shown as red filled circles. The flare start date is indicated by the dashed vertical line. The left axis indicates the units for the VHE flux and the right (red) axis denotes the units for the HE flux. (Bottom) The spectral energy distribution of the HE-VHE emission. For the HE emission the overall flare spectrum is shown as reported by \citet{psrb1259_fermi}. Marking of the data points is the same as in the (Top). Solid lines denote the fit of the Fermi data alone with the power law with exponential cut-off (red) and the fit of the \hess\ data alone with the power law (black). The dashed black line denotes the fit of the Fermi (excluding upper limits) and H.E.S.S. data together with the power law.}
  \label{pref_vs_flare_fig}
  \end{figure}

In order to search for variability, the flux as a function of time was fitted with 
a constant resulting in a mean flux of $(0.91\pm0.18)\times10^{-12}$ \fluxunits\ (black 
horizontal dashed line in Fig. 
\ref{pref_vs_flare_fig} (Top)). The fit has a 
$\chi^2$ to NDF ratio of $0.73/1$ which corresponds to a $\chi^2$ 
probability of 0.39, showing 
no indication for a flux enhancement. Note that the spectral 
parameters obtained by an 
independent fit of each of the two periods have been used here.

If one assumes that HE and VHE emission are created in the 
same scenario, i.e. the
same acceleration and radiation processes and sites, 
then a flux enhancement of the same magnitude as observed at HEs should be also seen 
at VHEs. In order to investigate this hypothesis, the flare coefficient $\kappa$ is 
introduced as the ratio of the fluxes during the flare period and the 
pre-flare period. The ratio of the HE ($E > 0.1$ GeV) flux averaged over 
the three days interval between $(t_{\mathrm{p}} + 30\,\mathrm{d})$ and 
$(t_{\mathrm{p}} + 32\,\mathrm{d})$ to the upper limit on the HE pre-flare emission (see Fig. 
\ref{pref_vs_flare_fig} (Top)) yields a lower limit on the HE emission flare 
coefficient $\kappa_{\mathrm{HE}} \geq 9.2$. An upper limit on the VHE flare 
coefficient can be estimated using the profile likelihood method. The likelihood 
function is defined as a product of two Gaussian distributions of the pre-flare and 
flare flux measurements $\phi_1$ and $\phi_2$ correspondingly, stating that the flare measurement 
$\phi_2$ varies around $\tilde{\kappa} \tilde{\phi_1}$, where the tilde denotes the true 
value for a parameter. The profile likelihood $\lambda$ is 
then built as a function of $\kappa$: 
\beq
\lambda(\kappa) = \frac{L(\hat{\phi}_{1}, \kappa | \phi_{1}, \phi_{2})}{L(\hat{\phi}_{1}, \hat{\kappa}| \phi_{1}, \phi_{2})}.
\enq
where the hat denotes the maximum likelihood estimate for a parameter. 
The variable $-2\log{\lambda}$ follows a $\chi^{2}$ distribution with 
1 degree of freedom, which 
allows to calculate the 99.7 \% confidence level (equivalent of 3$\sigma$) 
upper limit of $\kappa_{99.7\%}<3.5$. The obtained upper limit is lower than 
the observed lower limit on $\kappa_{\mathrm{HE}}$.     

The statistical tests presented above give two main results:
\begin{itemize}
\item A flare of similar magnitude as observed in the HE band can 
be firmly rejected in the VHE band at the same orbital phase.
\item There is no significant difference between the pre-flare and 
the flare flux in the VHE band.
\end{itemize}
These two results suggest that the HE flare emission has a 
different nature than the VHE emission.

This conclusion is also further supported by the inconsistency 
of HE and VHE emission spectra (see Fig. \ref{pref_vs_flare_fig} (Bottom)). 
The joint fit of the Fermi and H.E.S.S. data points with the simple 
power law (the dashed line on  Fig. \ref{pref_vs_flare_fig} (Bottom)) 
results in a fit probability of 0.004 and, hence, fails to explain 
the combined HE/VHE emission even ignoring Fermi upper limits which 
cannot be taken into account in the fit procedure. Moreover, the Fermi 
upper limits at $1-100$ GeV violate any reasonable model to explain the HE and 
VHE emission together. The Fermi spectrum alone is consistent with the 
the power law with exponential cut-off $E^2\mathrm{d}N/\mathrm{d}E = 
N_{0} \left( E/0.1 \mathrm{GeV} \right)^{-p} e^{-E/E_{\mathrm{cutoff}}}$ with 
the index $p = 0.16 \pm 0.32$, the cut-off energy of 
$E_{\mathrm{cutoff}} = 0.5\pm0.2$ GeV and the normalisation 
$N_{0} = (4\pm0.4)\times 10^{-10}$ erg cm$^{-2}$ s$^{-1}$. 
The fit probability is 0.27.

Several models have been proposed to explain the VHE emission 
from the source. In a hadronic scenario, the VHE \gammaray\ emission 
could be produced in the interaction 
of the ultrarelativistic pulsar wind particles with the dense equatorial disk 
outflow with subsequent production of $\pi^{0}$ pions and hence VHE 
$\gamma$-rays \citep{kawachi2004, neronov2007}. However, the detection of the source 
before the expected disk passage in 2007 put the hadronic scenario under 
doubt, suggesting that the VHE emission should be created at least partly by 
leptonic processes \citep{psrb1259_hess09}. Within the leptonic scenario, 
VHE emission from \psrb\ is explained by the inverse Compton (IC) scattering of 
shock-accelerated electrons on stellar photons \citep{tavani_psrb1259, kirk99, 
dubus2006, khangulyan2007}.

Few possible explanations for the nature of the 
HE flare are discussed in the literature. One of them suggested by 
\citet{khangulyan2011_post} is based 
on IC scattering of the unshocked 
pulsar wind on the stellar and circumstellar disk photons. 
While the pulsar is inside the disk the IC scattering 
of the unshocked pulsar wind is supressed due to the high ram pressure. 
But right after the pulsar escapes the disk in the post-periastron phase 
the unshocked pulsar wind zone towards the observer increases significantly, 
while the density of the circumstellar disk photons is still high enough for 
the efficient IC scattering. Therefore, the enhancement of the HE flux 
is observed. This is not expected in the pre-periastron phase because the 
termination shock should expand towards the direction opposite to the observer. 
This model also predicts the difference between the HE 
and VHE emission, the latter expected to result 
from the upscattering of the stellar photons by the electrons 
accelerated at the termination shock between pulsar and stellar winds. 
Another explanation of the HE flare can 
be the Doppler boosting of the radiation created by 
the shocked pulsar wind \citep{bogovalov_2008, dubus2010, kong2012}. 
It is unclear though why the flare 
is not detected at other wavebands, since the Doppler boosting 
should enhance also X-ray and VHE \gammaray\ fluxes. 
This issue, however, can be resolved assuming a 
specific anisotropy of the pulsar wind and the difference 
of the emission behavior in different regions of the 
termination shock, the isotropic emission in the apex 
and the beamed emission in the tail of the shock \citep{kong2012}. 
In this particular case the HE flare is explained by the Doppler 
boosting of the synchrotron emission. 
\citet{psrb1259_fermi} suggested that the flare can also be 
explained by an anisotropy of the pulsar wind and/or 
stellar material. The anisotropy of electrons 
with the highest energies would cause an anisotropy of the synchrotron 
radiation at high energies. 
In this interpretation, the HE emission 
is produced by the synchrotron mechanism. The local increase of the stellar 
wind density would increase the Bremsstrahlung component which may also 
cause the HE flare. Regardless of which mechanism is 
responsible for the HE flare, the fact that it is observed only after 
the periastron indicates either a strong dependency of the HE emission on 
the geometry of the system, i.e. its configuration with respect to the direction 
to the observer, or some local perturbation of the stellar material. 
A detailed interpretation of the HE-VHE emission is beyond the scope 
of this paper and will be discussed in the joint MWL paper.

\section{Summary}
\label{summary}

The binary system \psrb\ was monitored by \hess\ around the periastron passage 
on 15th of December 2010. The observed flux and spectral shape are in good agreement with what 
was measured during previous periastron passages. The observations were 
performed at similar orbital phases as around the 2004 periastron passage, 
strengthening the evidence for 
the repetitive behaviour of the source at VHEs.

H.E.S.S. observations were part of a joint MWL campaign including also 
radio, optical, X-ray and HE observations. A spectacular flare observed 
at HEs with Fermi LAT overlapped in time with the H.E.S.S. observations. 
A careful statistical study showed that the HE flare does not have a 
counterpart at VHEs, indicating that the HE and VHE emissions are produced 
in different physical scenarios.

\begin{acknowledgements}

The support of the Namibian authorities and of the University of
Namibia in facilitating the construction and operation of H.E.S.S.\ is
gratefully acknowledged, as is the support by the German Ministry for
Education and Research (BMBF), the Max Planck Society, the French
Ministry for Research, the CNRS-IN2P3 and the Astroparticle
Interdisciplinary Programme of the CNRS, the U.K. Particle Physics and
Astronomy Research Council (PPARC), the IPNP of the Charles
University, the South African Department of Science and Technology and
National Research Foundation, and by the University of Namibia. We
appreciate the excellent work of the technical support staff in
Berlin, Durham, Hamburg, Heidelberg, Palaiseau, Paris, Saclay, and in
Namibia in the construction and operation of the equipment.
\end{acknowledgements}

\bibliographystyle{aa}
\bibliography{psrb1259_HESS_aa_final.bbl}

\begin{thebibliography}{40}
\expandafter\ifx\csname natexlab\endcsname\relax\def\natexlab#1{#1}\fi

\bibitem[{{Abdo} {et~al.}(2010){Abdo}, {Ackermann}, {Ajello}, {Allafort},
  {Antolini}, {Atwood}, {Axelsson}, {Baldini}, {Ballet}, {Barbiellini}, \&
  et~al.}]{Fermi_catalog1}
{Abdo}, A.~A., {Ackermann}, M., {Ajello}, M., {et~al.} 2010, \apjs, 188, 405

\bibitem[{{Abdo} {et~al.}(2011){Abdo}, {Ackermann}, {Ajello}, {Allafort},
  {Ballet}, {Barbiellini}, {Bastieri}, {Bechtol}, {Bellazzini}, {Berenji},
  {Blandford}, {Bonamente}, {Borgland}, {Bregeon}, {Brigida}, {Bruel},
  {Buehler}, {Buson}, {Caliandro}, {Cameron}, {Camilo}, {Caraveo}, {Cecchi},
  {Charles}, {Chaty}, {Chekhtman}, {Chernyakova}, {Cheung}, {Chiang},
  {Ciprini}, {Claus}, {Cohen-Tanugi}, {Cominsky}, {Corbel}, {Cutini},
  {D'Ammando}, {de Angelis}, {den Hartog}, {de Palma}, {Dermer}, {Digel},
  {Silva}, {Dormody}, {Drell}, {Drlica-Wagner}, {Dubois}, {Dubus}, {Dumora},
  {Enoto}, {Espinoza}, {Favuzzi}, {Fegan}, {Ferrara}, {Focke}, {Fortin},
  {Fukazawa}, {Funk}, {Fusco}, {Gargano}, {Gasparrini}, {Gehrels}, {Germani},
  {Giglietto}, {Giommi}, {Giordano}, {Giroletti}, {Glanzman}, {Godfrey},
  {Grenier}, {Grondin}, {Grove}, {Grundstrom}, {Guiriec}, {Gwon}, {Hadasch},
  {Harding}, {Hayashida}, {Hays}, {J{\'o}hannesson}, {Johnson}, {Johnson},
  {Johnston}, {Kamae}, {Katagiri}, {Kataoka}, {Keith}, {Kerr},
  {Kn{\"o}dlseder}, {Kramer}, {Kuss}, {Lande}, {Lee}, {Lemoine-Goumard},
  {Longo}, {Loparco}, {Lovellette}, {Lubrano}, {Manchester}, {Marelli},
  {Mazziotta}, {Michelson}, {Mitthumsiri}, {Mizuno}, {Moiseev}, {Monte},
  {Monzani}, {Morselli}, {Moskalenko}, {Murgia}, {Nakamori}, {Naumann-Godo},
  {Neronov}, {Nolan}, {Norris}, {Noutsos}, {Nuss}, {Ohsugi}, {Okumura},
  {Omodei}, {Orlando}, {Paneque}, {Parent}, {Pesce-Rollins}, {Pierbattista},
  {Piron}, {Porter}, {Possenti}, {Rain{\`o}}, {Rando}, {Ray}, {Razzano},
  {Razzaque}, {Reimer}, {Reimer}, {Reposeur}, {Ritz}, {Sadrozinski}, {Scargle},
  {Sgr{\`o}}, {Shannon}, {Siskind}, {Smith}, {Spandre}, {Spinelli},
  {Strickman}, {Suson}, {Takahashi}, {Tanaka}, {Thayer}, {Thayer}, {Thompson},
  {Thorsett}, {Tibaldo}, {Tibolla}, {Torres}, {Tosti}, {Troja}, {Uchiyama},
  {Usher}, {Vandenbroucke}, {Vasileiou}, {Vianello}, {Vitale}, {Waite}, {Wang},
  {Winer}, {Wolff}, {Wood}, {Wood}, {Yang}, {Ziegler}, \&
  {Zimmer}}]{psrb1259_fermi}
{Abdo}, A.~A., {Ackermann}, M., {Ajello}, M., {et~al.} 2011, \apjl, 736, L11

\bibitem[{{Aharonian} {et~al.}(2009){Aharonian}, {Akhperjanian}, {Anton},
  {Barres de Almeida}, {Bazer-Bachi}, {Becherini}, {Behera}, {Bernl{\"o}hr},
  {Bochow}, {Boisson}, {Bolmont}, {Borrel}, {Brucker}, {Brun}, {Brun},
  {B{\"u}hler}, {Bulik}, {B{\"u}sching}, {Boutelier}, {Chadwick},
  {Charbonnier}, {Chaves}, {Cheesebrough}, {Chounet}, {Clapson}, {Coignet},
  {Dalton}, {Daniel}, {Davids}, {Degrange}, {Deil}, {Dickinson},
  {Djannati-Ata{\"i}}, {Domainko}, {O'C.~Drury}, {Dubois}, {Dubus}, {Dyks},
  {Dyrda}, {Egberts}, {Emmanoulopoulos}, {Espigat}, {Farnier}, {Feinstein},
  {Fiasson}, {F{\"o}rster}, {Fontaine}, {F{\"u}{\ss}ling}, {Gabici}, {Gallant},
  {G{\'e}rard}, {Gerbig}, {Giebels}, {Glicenstein}, {Gl{\"u}ck}, {Goret},
  {G{\"o}ring}, {Hauser}, {Hauser}, {Heinz}, {Heinzelmann}, {Henri}, {Hermann},
  {Hinton}, {Hoffmann}, {Hofmann}, {Holleran}, {Hoppe}, {Horns},
  {Jacholkowska}, {de Jager}, {Jahn}, {Jung}, {Katarzy{\'n}ski}, {Katz},
  {Kaufmann}, {Kerschhaggl}, {Khangulyan}, {Kh{\'e}lifi}, {Keogh}, {Klochkov},
  {Klu{\'z}niak}, {Kneiske}, {Komin}, {Kosack}, {Kossakowski}, {Lamanna},
  {Lenain}, {Lohse}, {Marandon}, {Martineau-Huynh}, {Marcowith}, {Masbou},
  {Maurin}, {McComb}, {Medina}, {Moderski}, {Moulin}, {Naumann-Godo}, {de
  Naurois}, {Nedbal}, {Nekrassov}, {Nicholas}, {Niemiec}, {Nolan}, {Ohm},
  {Olive}, {de O{\~n}a Wilhelmi}, {Orford}, {Ostrowski}, {Panter}, {Paz
  Arribas}, {Pedaletti}, {Pelletier}, {Petrucci}, {Pita}, {P{\"u}hlhofer},
  {Punch}, {Quirrenbach}, {Raubenheimer}, {Raue}, {Rayner}, {Renaud}, {Rieger},
  {Ripken}, {Rob}, {Rosier-Lees}, {Rowell}, {Rudak}, {Rulten}, {Ruppel},
  {Sahakian}, {Santangelo}, {Schlickeiser}, {Sch{\"o}ck}, {Schwanke},
  {Schwarzburg}, {Schwemmer}, {Shalchi}, {Sikora}, {Skilton}, {Sol},
  {Spangler}, {Stawarz}, {Steenkamp}, {Stegmann}, {Stinzing}, {Superina},
  {Szostek}, {Tam}, {Tavernet}, {Terrier}, {Tibolla}, {Tluczykont}, {van
  Eldik}, {Vasileiadis}, {Venter}, {Venter}, {Vialle}, {Vincent}, {Vivier},
  {V{\"o}lk}, {Volpe}, {Wagner}, {Ward}, {Zdziarski}, \&
  {Zech}}]{psrb1259_hess09}
{Aharonian}, F., {Akhperjanian}, A.~G., {Anton}, G., {et~al.} 2009, \aap, 507,
  389

\bibitem[{{Aharonian} {et~al.}(2005{\natexlab{a}}){Aharonian}, {Akhperjanian},
  {Aye}, {Bazer-Bachi}, {Beilicke}, {Benbow}, {Berge}, {Berghaus},
  {Bernl{\"o}hr}, {Boisson}, {Bolz}, {Borrel}, {Braun}, {Breitling}, {Brown},
  {Gordo}, {Chadwick}, {Chounet}, {Cornils}, {Costamante}, {Degrange},
  {Dickinson}, {Djannati-Ata{\"i}}, {Drury}, {Dubus}, {Emmanoulopoulos},
  {Espigat}, {Feinstein}, {Fleury}, {Fontaine}, {Fuchs}, {Funk}, {Gallant},
  {Giebels}, {Gillessen}, {Glicenstein}, {Goret}, {Hadjichristidis}, {Hauser},
  {Heinzelmann}, {Henri}, {Hermann}, {Hinton}, {Hofmann}, {Holleran}, {Horns},
  {Jacholkowska}, {de Jager}, {Kh{\'e}lifi}, {Komin}, {Konopelko}, {Latham},
  {Le Gallou}, {Lemi{\`e}re}, {Lemoine-Goumard}, {Leroy}, {Lohse}, {Marcowith},
  {Martin}, {Martineau-Huynh}, {Masterson}, {McComb}, {de Naurois}, {Nolan},
  {Noutsos}, {Orford}, {Osborne}, {Ouchrif}, {Panter}, {Pelletier}, {Pita},
  {P{\"u}hlhofer}, {Punch}, {Raubenheimer}, {Raue}, {Raux}, {Rayner}, {Reimer},
  {Reimer}, {Ripken}, {Rob}, {Rolland}, {Rowell}, {Sahakian}, {Saug{\'e}},
  {Schlenker}, {Schlickeiser}, {Schuster}, {Schwanke}, {Siewert}, {Sol},
  {Spangler}, {Steenkamp}, {Stegmann}, {Tavernet}, {Terrier}, {Th{\'e}oret},
  {Tluczykont}, {Vasileiadis}, {Venter}, {Vincent}, {V{\"o}lk}, \&
  {Wagner}}]{LS_5039}
{Aharonian}, F., {Akhperjanian}, A.~G., {Aye}, K.-M., {et~al.}
  2005{\natexlab{a}}, Science, 309, 746

\bibitem[{{Aharonian} {et~al.}(2005{\natexlab{b}}){Aharonian}, {Akhperjanian},
  {Aye}, {Bazer-Bachi}, {Beilicke}, {Benbow}, {Berge}, {Berghaus},
  {Bernl{\"o}hr}, {Boisson}, {Bolz}, {Braun}, {Breitling}, {Brown}, {Bussons
  Gordo}, {Chadwick}, {Chounet}, {Cornils}, {Costamante}, {Degrange},
  {Djannati-Ata{\"i}}, {O'C.~Drury}, {Dubus}, {Emmanoulopoulos}, {Espigat},
  {Feinstein}, {Fleury}, {Fontaine}, {Fuchs}, {Funk}, {Gallant}, {Giebels},
  {Gillessen}, {Glicenstein}, {Goret}, {Hadjichristidis}, {Hauser},
  {Heinzelmann}, {Henri}, {Hermann}, {Hinton}, {Hofmann}, {Holleran}, {Horns},
  {de Jager}, {Johnston}, {Kh{\'e}lifi}, {Kirk}, {Komin}, {Konopelko},
  {Latham}, {Le Gallou}, {Lemi{\`e}re}, {Lemoine-Goumard}, {Leroy},
  {Martineau-Huynh}, {Lohse}, {Marcowith}, {Masterson}, {McComb}, {de Naurois},
  {Nolan}, {Noutsos}, {Orford}, {Osborne}, {Ouchrif}, {Panter}, {Pelletier},
  {Pita}, {P{\"u}hlhofer}, {Punch}, {Raubenheimer}, {Raue}, {Raux}, {Rayner},
  {Redondo}, {Reimer}, {Reimer}, {Ripken}, {Rob}, {Rolland}, {Rowell},
  {Sahakian}, {Saug{\'e}}, {Schlenker}, {Schlickeiser}, {Schuster}, {Schwanke},
  {Siewert}, {Skj{\ae}raasen}, {Sol}, {Steenkamp}, {Stegmann}, {Tavernet},
  {Terrier}, {Th{\'e}oret}, {Tluczykont}, {Vasileiadis}, {Venter}, {Vincent},
  {V{\"o}lk}, \& {Wagner}}]{psrb1259_hess05}
{Aharonian}, F., {Akhperjanian}, A.~G., {Aye}, K.-M., {et~al.}
  2005{\natexlab{b}}, \aap, 442, 1

\bibitem[{{Aharonian} {et~al.}(2006){Aharonian}, {Akhperjanian}, {Bazer-Bachi},
  {Beilicke}, {Benbow}, {Berge}, {Bernl{\"o}hr}, {Boisson}, {Bolz}, {Borrel},
  {Braun}, {Breitling}, {Brown}, {B{\"u}hler}, {B{\"u}sching}, {Carrigan},
  {Chadwick}, {Chounet}, {Cornils}, {Costamante}, {Degrange}, {Dickinson},
  {Djannati-Ata{\"i}}, {O'C.~Drury}, {Dubus}, {Egberts}, {Emmanoulopoulos},
  {Espigat}, {Feinstein}, {Ferrero}, {Fiasson}, {Fontaine}, {Funk}, {Funk},
  {Gallant}, {Giebels}, {Glicenstein}, {Goret}, {Hadjichristidis}, {Hauser},
  {Hauser}, {Heinzelmann}, {Henri}, {Hermann}, {Hinton}, {Hofmann}, {Holleran},
  {Horns}, {Jacholkowska}, {de Jager}, {Kh{\'e}lifi}, {Komin}, {Konopelko},
  {Kosack}, {Latham}, {Le Gallou}, {Lemi{\`e}re}, {Lemoine-Goumard}, {Lohse},
  {Martin}, {Martineau-Huynh}, {Marcowith}, {Masterson}, {McComb}, {de
  Naurois}, {Nedbal}, {Nolan}, {Noutsos}, {Orford}, {Osborne}, {Ouchrif},
  {Panter}, {Pelletier}, {Pita}, {P{\"u}hlhofer}, {Punch}, {Raubenheimer},
  {Raue}, {Rayner}, {Reimer}, {Reimer}, {Ripken}, {Rob}, {Rolland}, {Rowell},
  {Sahakian}, {Saug{\'e}}, {Schlenker}, {Schlickeiser}, {Schwanke}, {Sol},
  {Spangler}, {Spanier}, {Steenkamp}, {Stegmann}, {Superina}, {Tavernet},
  {Terrier}, {Th{\'e}oret}, {Tluczykont}, {van Eldik}, {Vasileiadis}, {Venter},
  {Vincent}, {V{\"o}lk}, {Wagner}, \& {Ward}}]{crab}
{Aharonian}, F., {Akhperjanian}, A.~G., {Bazer-Bachi}, A.~R., {et~al.} 2006,
  \aap, 457, 899

\bibitem[{{Aharonian} {et~al.}(2007){Aharonian}, {Akhperjanian}, {Bazer-Bachi},
  {Behera}, {Beilicke}, {Benbow}, {Berge}, {Bernl{\"o}hr}, {Boisson}, {Bolz},
  {Borrel}, {Braun}, {Brion}, {Brown}, {B{\"u}hler}, {B{\"u}sching},
  {Boutelier}, {Carrigan}, {Chadwick}, {Chounet}, {Coignet}, {Cornils},
  {Costamante}, {Degrange}, {Dickinson}, {Djannati-Ata{\"i}}, {Domainko},
  {O'C.~Drury}, {Dubus}, {Egberts}, {Emmanoulopoulos}, {Espigat}, {Farnier},
  {Feinstein}, {Fiasson}, {F{\"o}rster}, {Fontaine}, {Funk}, {Funk},
  {F{\"u}{\ss}ling}, {Gallant}, {Giebels}, {Glicenstein}, {Gl{\"u}ck}, {Goret},
  {Hadjichristidis}, {Hauser}, {Hauser}, {Heinzelmann}, {Henri}, {Hermann},
  {Hinton}, {Hoffmann}, {Hofmann}, {Holleran}, {Hoppe}, {Horns},
  {Jacholkowska}, {de Jager}, {Kendziorra}, {Kerschhaggl}, {Kh{\'e}lifi},
  {Komin}, {Kosack}, {Lamanna}, {Latham}, {Le Gallou}, {Lemi{\`e}re},
  {Lemoine-Goumard}, {Lohse}, {Martin}, {Martineau-Huynh}, {Marcowith},
  {Masterson}, {Maurin}, {McComb}, {Moulin}, {de Naurois}, {Nedbal}, {Nolan},
  {Noutsos}, {Olive}, {Orford}, {Osborne}, {Panter}, {Pedaletti}, {Pelletier},
  {Petrucci}, {Pita}, {P{\"u}hlhofer}, {Punch}, {Ranchon}, {Raubenheimer},
  {Raue}, {Rayner}, {Reimer}, {Ripken}, {Rob}, {Rolland}, {Rosier-Lees},
  {Rowell}, {Ruppel}, {Sahakian}, {Santangelo}, {Saug{\'e}}, {Schlenker},
  {Schlickeiser}, {Schr{\"o}der}, {Schwanke}, {Schwarzburg}, {Schwemmer},
  {Shalchi}, {Sol}, {Spangler}, {Steenkamp}, {Stegmann}, {Superina}, {Tam},
  {Tavernet}, {Terrier}, {Tluczykont}, {van Eldik}, {Vasileiadis}, {Venter},
  {Vialle}, {Vincent}, {V{\"o}lk}, {Wagner}, {Ward}, {Moriguchi}, \&
  {Fukui}}]{hess_j0632}
{Aharonian}, F.~A., {Akhperjanian}, A.~G., {Bazer-Bachi}, A.~R., {et~al.} 2007,
  \aap, 469, L1

\bibitem[{{Albert} {et~al.}(2006){Albert}, {Aliu}, {Anderhub}, {Antoranz},
  {Armada}, {Asensio}, {Baixeras}, {Barrio}, {Bartelt}, {Bartko}, {Bastieri},
  {Bavikadi}, {Bednarek}, {Berger}, {Bigongiari}, {Biland}, {Bisesi}, {Bock},
  {Bordas}, {Bosch-Ramon}, {Bretz}, {Britvitch}, {Camara}, {Carmona},
  {Chilingarian}, {Ciprini}, {Coarasa}, {Commichau}, {Contreras}, {Cortina},
  {Curtef}, {Danielyan}, {Dazzi}, {De Angelis}, {de los Reyes}, {De Lotto},
  {Domingo-Santamar{\'{\i}}a}, {Dorner}, {Doro}, {Errando}, {Fagiolini},
  {Ferenc}, {Fern{\'a}ndez}, {Firpo}, {Flix}, {Fonseca}, {Font}, {Fuchs},
  {Galante}, {Garczarczyk}, {Gaug}, {Giller}, {Goebel}, {Hakobyan},
  {Hayashida}, {Hengstebeck}, {H{\"o}hne}, {Hose}, {Hsu}, {Isar}, {Jacon},
  {Kalekin}, {Kosyra}, {Kranich}, {Laatiaoui}, {Laille}, {Lenisa}, {Liebing},
  {Lindfors}, {Lombardi}, {Longo}, {L{\'o}pez}, {L{\'o}pez}, {Lorenz},
  {Lucarelli}, {Majumdar}, {Maneva}, {Mannheim}, {Mansutti}, {Mariotti},
  {Mart{\'{\i}}nez}, {Mase}, {Mazin}, {Merck}, {Meucci}, {Meyer}, {Miranda},
  {Mirzoyan}, {Mizobuchi}, {Moralejo}, {Nilsson}, {O{\~n}a-Wilhelmi},
  {Ordu{\~n}a}, {Otte}, {Oya}, {Paneque}, {Paoletti}, {Paredes}, {Pasanen},
  {Pascoli}, {Pauss}, {Pavel}, {Pegna}, {Persic}, {Peruzzo}, {Piccioli},
  {Poller}, {Pooley}, {Prandini}, {Raymers}, {Rhode}, {Rib{\'o}}, {Rico},
  {Riegel}, {Rissi}, {Robert}, {Romero}, {R{\"u}gamer}, {Saggion},
  {S{\'a}nchez}, {Sartori}, {Scalzotto}, {Scapin}, {Schmitt}, {Schweizer},
  {Shayduk}, {Shinozaki}, {Shore}, {Sidro}, {Sillanp{\"a}{\"a}}, {Sobczynska},
  {Stamerra}, {Stark}, {Takalo}, {Temnikov}, {Tescaro}, {Teshima}, {Tonello},
  {Torres}, {Torres}, {Turini}, {Vankov}, {Vitale}, {Wagner}, {Wibig},
  {Wittek}, {Zanin}, \& {Zapatero}}]{LS_61303}
{Albert}, J., {Aliu}, E., {Anderhub}, H., {et~al.} 2006, Science, 312, 1771

\bibitem[{{Albert} {et~al.}(2007){Albert}, {Aliu}, {Anderhub}, {Antoranz},
  {Armada}, {Baixeras}, {Barrio}, {Bartko}, {Bastieri}, {Becker}, {Bednarek},
  {Berger}, {Bigongiari}, {Biland}, {Bock}, {Bordas}, {Bosch-Ramon}, {Bretz},
  {Britvitch}, {Camara}, {Carmona}, {Chilingarian}, {Coarasa}, {Commichau},
  {Contreras}, {Cortina}, {Costado}, {Curtef}, {Danielyan}, {Dazzi}, {De
  Angelis}, {Delgado}, {de los Reyes}, {De Lotto}, {Domingo-Santamar{\'{\i}}a},
  {Dorner}, {Doro}, {Errando}, {Fagiolini}, {Ferenc}, {Fern{\'a}ndez}, {Firpo},
  {Flix}, {Fonseca}, {Font}, {Fuchs}, {Galante}, {Garc{\'{\i}}a-L{\'o}pez},
  {Garczarczyk}, {Gaug}, {Giller}, {Goebel}, {Hakobyan}, {Hayashida},
  {Hengstebeck}, {Herrero}, {H{\"o}hne}, {Hose}, {Hsu}, {Jacon}, {Jogler},
  {Kosyra}, {Kranich}, {Kritzer}, {Laille}, {Lindfors}, {Lombardi}, {Longo},
  {L{\'o}pez}, {L{\'o}pez}, {Lorenz}, {Majumdar}, {Maneva}, {Mannheim},
  {Mansutti}, {Mariotti}, {Mart{\'{\i}}nez}, {Mazin}, {Merck}, {Meucci},
  {Meyer}, {Miranda}, {Mirzoyan}, {Mizobuchi}, {Moralejo}, {Nieto}, {Nilsson},
  {Ninkovic}, {O{\~n}a-Wilhelmi}, {Otte}, {Oya}, {Panniello}, {Paoletti},
  {Paredes}, {Pasanen}, {Pascoli}, {Pauss}, {Pegna}, {Persic}, {Peruzzo},
  {Piccioli}, {Prandini}, {Puchades}, {Raymers}, {Rhode}, {Rib{\'o}}, {Rico},
  {Rissi}, {Robert}, {R{\"u}gamer}, {Saggion}, {Saito}, {S{\'a}nchez},
  {Sartori}, {Scalzotto}, {Scapin}, {Schmitt}, {Schweizer}, {Shayduk},
  {Shinozaki}, {Shore}, {Sidro}, {Sillanp{\"a}{\"a}}, {Sobczynska}, {Stamerra},
  {Stark}, {Takalo}, {Temnikov}, {Tescaro}, {Teshima}, {Torres}, {Turini},
  {Vankov}, {Vitale}, {Wagner}, {Wibig}, {Wittek}, {Zandanel}, {Zanin}, \&
  {Zapatero}}]{cygnusX}
{Albert}, J., {Aliu}, E., {Anderhub}, H., {et~al.} 2007, \apjl, 665, L51

\bibitem[{{Bogovalov} {et~al.}(2008){Bogovalov}, {Khangulyan}, {Koldoba},
  {Ustyugova}, \& {Aharonian}}]{bogovalov_2008}
{Bogovalov}, S.~V., {Khangulyan}, D.~V., {Koldoba}, A.~V., {Ustyugova}, G.~V.,
  \& {Aharonian}, F.~A. 2008, \mnras, 387, 63

\bibitem[{{Chernyakova} {et~al.}(2009){Chernyakova}, {Neronov}, {Aharonian},
  {Uchiyama}, \& {Takahashi}}]{chernyakova2009}
{Chernyakova}, M., {Neronov}, A., {Aharonian}, F., {Uchiyama}, Y., \&
  {Takahashi}, T. 2009, \mnras, 397, 2123

\bibitem[{{Chernyakova} {et~al.}(2006){Chernyakova}, {Neronov}, {Lutovinov},
  {Rodriguez}, \& {Johnston}}]{chernyakova2006}
{Chernyakova}, M., {Neronov}, A., {Lutovinov}, A., {Rodriguez}, J., \&
  {Johnston}, S. 2006, \mnras, 367, 1201

\bibitem[{{Cominsky} {et~al.}(1994){Cominsky}, {Roberts}, \&
  {Johnston}}]{cominsky94}
{Cominsky}, L., {Roberts}, M., \& {Johnston}, S. 1994, \apj, 427, 978

\bibitem[{{Connors} {et~al.}(2002){Connors}, {Johnston}, {Manchester}, \&
  {McConnell}}]{connors2002}
{Connors}, T.~W., {Johnston}, S., {Manchester}, R.~N., \& {McConnell}, D. 2002,
  \mnras, 336, 1201

\bibitem[{{de Naurois} \& {Rolland}(2009)}]{deNaurois09}
{de Naurois}, M. \& {Rolland}, L. 2009, Astroparticle Physics, 32, 231

\bibitem[{{Dubus}(2006)}]{dubus2006}
{Dubus}, G. 2006, \aap, 456, 801

\bibitem[{{Dubus} {et~al.}(2010){Dubus}, {Cerutti}, \& {Henri}}]{dubus2010}
{Dubus}, G., {Cerutti}, B., \& {Henri}, G. 2010, \aap, 516, A18

\bibitem[{{HESS Collaboration} {et~al.}(2012){HESS Collaboration},
  {Abramowski}, {Acero}, {Aharonian}, {Akhperjanian}, {Anton}, {Balzer},
  {Barnacka}, {Barres de Almeida}, {Becherini}, {Becker}, {Behera},
  {Bernl{\"o}hr}, {Birsin}, {Biteau}, {Bochow}, {Boisson}, {Bolmont}, {Bordas},
  {Brucker}, {Brun}, {Brun}, {Bulik}, {B{\"u}sching}, {Carrigan}, {Casanova},
  {Cerruti}, {Chadwick}, {Charbonnier}, {Chaves}, {Cheesebrough}, {Clapson},
  {Coignet}, {Cologna}, {Conrad}, {Dalton}, {Daniel}, {Davids}, {Degrange},
  {Deil}, {Dickinson}, {Djannati-Ata{\"i}}, {Domainko}, {O'C.~Drury}, {Dubus},
  {Dutson}, {Dyks}, {Dyrda}, {Egberts}, {Eger}, {Espigat}, {Fallon}, {Farnier},
  {Fegan}, {Feinstein}, {Fernandes}, {Fiasson}, {Fontaine}, {F{\"o}rster},
  {F{\"u}{\ss}ling}, {Gallant}, {Gast}, {G{\'e}rard}, {Gerbig}, {Giebels},
  {Glicenstein}, {Gl{\"u}ck}, {Goret}, {G{\"o}ring}, {H{\"a}ffner}, {Hague},
  {Hampf}, {Hauser}, {Heinz}, {Heinzelmann}, {Henri}, {Hermann}, {Hinton},
  {Hoffmann}, {Hofmann}, {Hofverberg}, {Holler}, {Horns}, {Jacholkowska}, {de
  Jager}, {Jahn}, {Jamrozy}, {Jung}, {Kastendieck}, {Katarzy{\'n}ski}, {Katz},
  {Kaufmann}, {Keogh}, {Khangulyan}, {Kh{\'e}lifi}, {Klochkov}, {Klu{\'z}niak},
  {Kneiske}, {Komin}, {Kosack}, {Kossakowski}, {Laffon}, {Lamanna}, {Lennarz},
  {Lohse}, {Lopatin}, {Lu}, {Marandon}, {Marcowith}, {Masbou}, {Maurin},
  {Maxted}, {Mayer}, {McComb}, {Medina}, {M{\'e}hault}, {Moderski}, {Moulin},
  {Naumann}, {Naumann-Godo}, {de Naurois}, {Nedbal}, {Nekrassov}, {Nguyen},
  {Nicholas}, {Niemiec}, {Nolan}, {Ohm}, {de O{\~n}a Wilhelmi}, {Opitz},
  {Ostrowski}, {Oya}, {Panter}, {Paz Arribas}, {Pedaletti1}, {Pelletier},
  {Petrucci}, {Pita}, {P{\"u}hlhofer}, {Punch}, {Quirrenbach}, {Raue},
  {Rayner}, {Reimer}, {Reimer}, {Renaud}, {de los Reyes}, {Rieger}, {Ripken},
  {Rob}, {Rosier-Lees}, {Rowell}, {Rudak}, {Rulten}, {Ruppel}, {Sahakian},
  {Sanchez}, {Santangelo}, {Schlickeiser}, {Sch{\"o}ck}, {Schulz}, {Schwanke},
  {Schwarzburg}, {Schwemmer}, {Sheidaei}, {Skilton}, {Sol}, {Spengler},
  {Stawarz}, {Steenkamp}, {Stegmann}, {Stinzing}, {Stycz}, {Sushch}, {Szostek},
  {Tavernet}, {Terrier}, {Tluczykont}, {Valerius}, {van Eldik}, {Vasileiadis},
  {Venter}, {Vialle}, {Viana}, {Vincent}, {V{\"o}lk}, {Volpe}, {Vorobiov},
  {Vorster}, {Wagner}, {Ward}, {White}, {Wierzcholska}, {Zacharias}, {Zajczyk},
  {Zdziarski}, {Zech}, \& {Zechlin}}]{hess_j1018}
{HESS Collaboration}, {Abramowski}, A., {Acero}, F., {et~al.} 2012, ArXiv
  e-prints:1203.3215

\bibitem[{{Hinton}(2004)}]{hinton_2004}
{Hinton}, J.~A. 2004, \nar, 48, 331

\bibitem[{{Hirayama} {et~al.}(1999){Hirayama}, {Cominsky}, {Kaspi}, {Nagase},
  {Tavani}, {Kawai}, \& {Grove}}]{hirayama99}
{Hirayama}, M., {Cominsky}, L.~R., {Kaspi}, V.~M., {et~al.} 1999, \apj, 521,
  718

\bibitem[{{Johnston} {et~al.}(2005){Johnston}, {Ball}, {Wang}, \&
  {Manchester}}]{johnston2005}
{Johnston}, S., {Ball}, L., {Wang}, N., \& {Manchester}, R.~N. 2005, \mnras,
  358, 1069

\bibitem[{{Johnston} {et~al.}(1992{\natexlab{a}}){Johnston}, {Lyne},
  {Manchester}, {Kniffen}, {D'Amico}, {Lim}, \& {Ashworth}}]{johnston1}
{Johnston}, S., {Lyne}, A.~G., {Manchester}, R.~N., {et~al.}
  1992{\natexlab{a}}, \mnras, 255, 401

\bibitem[{{Johnston} {et~al.}(1992{\natexlab{b}}){Johnston}, {Manchester},
  {Lyne}, {Bailes}, {Kaspi}, {Qiao}, \& {D'Amico}}]{johnston2}
{Johnston}, S., {Manchester}, R.~N., {Lyne}, A.~G., {et~al.}
  1992{\natexlab{b}}, \apjl, 387, L37

\bibitem[{{Johnston} {et~al.}(1999){Johnston}, {Manchester}, {McConnell}, \&
  {Campbell-Wilson}}]{johnston99}
{Johnston}, S., {Manchester}, R.~N., {McConnell}, D., \& {Campbell-Wilson}, D.
  1999, \mnras, 302, 277

\bibitem[{{Kaspi} {et~al.}(1995){Kaspi}, {Tavani}, {Nagase}, {Hirayama},
  {Hoshino}, {Aoki}, {Kawai}, \& {Arons}}]{kaspi95}
{Kaspi}, V.~M., {Tavani}, M., {Nagase}, F., {et~al.} 1995, \apj, 453, 424

\bibitem[{{Kawachi} {et~al.}(2004){Kawachi}, {Naito}, {Patterson}, {Edwards},
  {Asahara}, {Bicknell}, {Clay}, {Enomoto}, {Gunji}, {Hara}, {Hara}, {Hattori},
  {Hayashi}, {Hayashi}, {Itoh}, {Kabuki}, {Kajino}, {Katagiri}, {Kifune},
  {Ksenofontov}, {Kubo}, {Kushida}, {Matsubara}, {Mizumoto}, {Mori}, {Moro},
  {Muraishi}, {Muraki}, {Nakase}, {Nishida}, {Nishijima}, {Ohishi}, {Okumura},
  {Protheroe}, {Sakurazawa}, {Swaby}, {Tanimori}, {Tokanai}, {Tsuchiya},
  {Tsunoo}, {Uchida}, {Watanabe}, {Watanabe}, {Yanagita}, {Yoshida}, \&
  {Yoshikoshi}}]{kawachi2004}
{Kawachi}, A., {Naito}, T., {Patterson}, J.~R., {et~al.} 2004, \apj, 607, 949

\bibitem[{{Kerschhaggl}(2011)}]{kerschhaggl_2011}
{Kerschhaggl}, M. 2011, \aap, 525, A80

\bibitem[{{Khangulyan} {et~al.}(2012){Khangulyan}, {Aharonian}, {Bogovalov}, \&
  {Rib{\'o}}}]{khangulyan2011_post}
{Khangulyan}, D., {Aharonian}, F.~A., {Bogovalov}, S.~V., \& {Rib{\'o}}, M.
  2012, \apjl, 752, L17

\bibitem[{{Khangulyan} {et~al.}(2007){Khangulyan}, {Hnatic}, {Aharonian}, \&
  {Bogovalov}}]{khangulyan2007}
{Khangulyan}, D., {Hnatic}, S., {Aharonian}, F., \& {Bogovalov}, S. 2007,
  \mnras, 380, 320

\bibitem[{{Kirk} {et~al.}(1999){Kirk}, {Ball}, \& {Skjaeraasen}}]{kirk99}
{Kirk}, J.~G., {Ball}, L., \& {Skjaeraasen}, O. 1999, Astroparticle Physics,
  10, 31

\bibitem[{{Kong} {et~al.}(2012){Kong}, {Cheng}, \& {Huang}}]{kong2012}
{Kong}, S.~W., {Cheng}, K.~S., \& {Huang}, Y.~F. 2012, \apj, 753, 127

\bibitem[{{Li} \& {Ma}(1983)}]{lima}
{Li}, T.-P. \& {Ma}, Y.-Q. 1983, \apj, 272, 317

\bibitem[{{Melatos} {et~al.}(1995){Melatos}, {Johnston}, \&
  {Melrose}}]{melatos}
{Melatos}, A., {Johnston}, S., \& {Melrose}, D.~B. 1995, \mnras, 275, 381

\bibitem[{{Mold{\'o}n} {et~al.}(2011){Mold{\'o}n}, {Johnston}, {Rib{\'o}},
  {Paredes}, \& {Deller}}]{moldon2011}
{Mold{\'o}n}, J., {Johnston}, S., {Rib{\'o}}, M., {Paredes}, J.~M., \&
  {Deller}, A.~T. 2011, \apjl, 732, L10

\bibitem[{{Negueruela} {et~al.}(2011){Negueruela}, {Rib{\'o}}, {Herrero},
  {Lorenzo}, {Khangulyan}, \& {Aharonian}}]{negueruela}
{Negueruela}, I., {Rib{\'o}}, M., {Herrero}, A., {et~al.} 2011, \apjl, 732, L11

\bibitem[{{Neronov} \& {Chernyakova}(2007)}]{neronov2007}
{Neronov}, A. \& {Chernyakova}, M. 2007, \apss, 309, 253

\bibitem[{{Shaw} {et~al.}(2004){Shaw}, {Chernyakova}, {Rodriguez}, {Walter},
  {Kretschmar}, \& {Mereghetti}}]{shaw2004}
{Shaw}, S.~E., {Chernyakova}, M., {Rodriguez}, J., {et~al.} 2004, \aap, 426,
  L33

\bibitem[{{Tam} {et~al.}(2011){Tam}, {Huang}, {Takata}, {Hui}, {Kong}, \&
  {Cheng}}]{tam}
{Tam}, P.~H.~T., {Huang}, R.~H.~H., {Takata}, J., {et~al.} 2011, \apjl, 736,
  L10

\bibitem[{{Tavani} \& {Arons}(1997)}]{tavani_psrb1259}
{Tavani}, M. \& {Arons}, J. 1997, \apj, 477, 439

\bibitem[{{Wang} {et~al.}(2004){Wang}, {Johnston}, \& {Manchester}}]{wang_2004}
{Wang}, N., {Johnston}, S., \& {Manchester}, R.~N. 2004, \mnras, 351, 599

\end{thebibliography}

\end{document}